\documentstyle[12pt]{article}
\def\lappeq{\mathrel{\rlap{\raise.5ex\hbox{$<$}}
{\lower.5ex\hbox{$\sim$}}}}
\def\beq{\begin{equation}}
\def\eeq{\end{equation}}
\begin{document}
\begin{titlepage}
\pagestyle{empty}
\baselineskip=21pt
\rightline{Alberta Thy-05-00}
\rightline{hep-ph/0004035}
\rightline{April 2000}
\vskip .05in
\begin{center}
{\large{\bf Thermalization After Inflation and Reheating Temperature}}
\end{center}
\vskip .05in
\begin{center}
{\bf Rouzbeh Allahverdi} 

{\it Department of Physics, University of Alberta}

{\it Edmonton, Alberta, Canada T6G 2J1}

\vskip .05in

\end{center}
\centerline{ {\bf Abstract} }
\baselineskip=18pt

\noindent

We present a detailed examination of thermalization after inflation for perturbative inflaton decay.  Different interactions among particles in the plasma of inflaton decay products are considered and it will be shown that  $2 \rightarrow 2$ scatterings and particle decay are the important ones.  We show that thermalization occurs after decays dominate scatterings, and that depending on the typical mass scale of inflaton decay products, different situations may arise.  In particular, thermalization may be delayed until late times, in which case the bounds from thermal gravitino production on supersymmetric models of inflation are considerably relaxed.  We will also consider the case where the observable sector consists only of the MSSM matter content, and point out that flat directions with large vevs may result in earlier thermalization of the plasma and push the reheat temperature towards its upper limit.      
  
 \vskip .1in

\end{titlepage}
\baselineskip=18pt
{\newcommand{\la}{\mbox{\raisebox{-.6ex}{$\stackrel{<}{\sim}$}}}
{\newcommand{\ga}{\mbox{\raisebox{-.6ex}{$\stackrel{>}{\sim}$}}}

\section{Introduction}

According to inflationary models \cite{infl}, which were first considered to address the flatness, isotropy, and (depending on the particle physics model of the early universe) monopole problems of the hot big-bang model, the universe has undergone several stages during its evolution.  During inflation, the energy density of the universe is dominated by the potential energy of the inflaton and the universe experiences a period of superluminal expansion. After inflation, the coherent oscillations of the inflaton, which behave like non-relativistic matter, dominate the energy density of the universe.  At some later time these coherent oscillations decay to the fields to which they are coupled, and their energy density is transferred to relativistic particles, the reheating stage, which results in a radiation-dominated FRW universe.  After the inflaton decay products are thermalized, the dynamics of the universe will be that of the hot big-bang model.  

Until recently, reheating was treated as a perturbative, one particle decay of the inflaton with the decay rate ${\Gamma}_{d}$ (depending on the microphysics), leading to the simple estimate ${T}_{R} \sim {({\Gamma}_{d}{M}_{Pl})}^{1 \over 2}$ for the reheat temperature (assuming instant thermalization) \cite{reheat}.  ${T}_{R}$ should be low enough so that the GUT symmetry is not restored and the original monopole problem is avoided.  In supersymmetric models there are even stricter bounds on the reheat temperature.  Gravitinos (the spin-${3 \over 2}$ superpartners of gravitons) with a mass in the range of 100 GeV-1 TeV (in agreement with low energy supersymmetry) decay after the big-bang nucleosynthesis.  They are also produced in a thermal bath, predominantly through $2 \rightarrow 2$ scatterings of gauge fields and gauginos.  This results in the bound ${T}_{R} {\ \lower-1.2pt\vbox{\hbox{\rlap{$<$}\lower5pt\vbox{\hbox{$\sim$}}}}\ } {10}^{8}-{10}^{10}$ GeV, in order to avoid gravitino overproduction which would destroy the successful predictions of big-bang nucleosynthesis \cite{gravitino}.  It has recently been noted that the initial stage of inflaton decay might occur through a complicated and non-perturbative process called parametric resonance, leading to an out-of-equilibrium distribution of final state particles with energies much higher than the inflaton mass (the preheating stage) \cite{preheat1,preheat2}.  This may have important cosmological implications, including gravitino overproduction which is most relevnt to our discussion.  Gravitinos can be produced either through the scattering of particles in the preheat distribution \cite{y}, or directly during inflaton oscillations\cite{lm,kklv,grt1,grt2}.  In particular, it has been recently noted that the second mechanism can lead to efficient production of helicity $\pm {1 \over 2}$ gravitinos, which possibly dominate the abundance of thermally produced gravitinos by several orders of magnitude \cite{kklv,grt1,grt2,l, bm}.   

However, it is believed that in a wide range of realistic inflationary models the final stage of inflaton decay occurs in the perturbative regime and that the reheat temperature is determined therein.  Thus a correct treatment of this stage and the subsequent thermalization of decay products is necessary.  After all, parametric resonance may not occur or may be modified, and a mechanism may be found for the dilution of relics which were produced in dangerous abundances during preheating.  In any case, the thermal production of such relics, gravitinos in particular, must be in accordance with nucleosynthesis bounds.

Our aim in this letter is to give a more detailed examination of thermalization after perturbative decay of the inflaton.  In this regime, inflaton decay products have energies higher than the mean thermal energy (if they were instantly thermalized) and a number density lower than the thermal equilibrium value.  This implies that thermalization occurs when the interactions which increase the number of particles are at equilibrium.  We will note that the inflaton decay is not sudden, but rather a prolonged process which starts very shortly after the end of inflation.  We then consider the energy distribution of particles in the plasma of inflaton decay products, and compare the rates for their decay and $2 \rightarrow N$ particle scatterings (with $N \geq 3$), as well as $2 \rightarrow 2$ particle scatterings which only redistribute the energy among the scattered particles.  It will be shown that (cascade) decay is the dominant process which leads to thermalization.  We find that, in general, a seed in the plasma of inflaton decay products is thermalized first and then the bulk of the plasma will be thermalized very rapidly by scattering off particles in the thermalized seed.  It is emphasized that the thermalization time depends on the the typical mass scale of those inflaton decay products which have interactions of gauge strength.  Finally, we will consider the case where the observable sector consists only of the MSSM matter content, and discuss the effect of flat direction vevs on thermalization.


\section{The Energy Spectrum of Inflaton Decay Products}

After inflation the inflaton starts its coherent oscillations around the minimum of the scalar potential, and at some later time it will decay to the fields to which it is coupled.  In the perturbative regime the decay occurs over many oscillations of the inflaton field.  This means that the oscillating inflaton field behaves like non-relativistic matter consisting of  a condensate of zero-mode bosons with the mass ${m}_{I}$.  The inflaton decay rate ${\Gamma}_{d}$ can then be calculated from the one particle decay channel of these bosons. The value of ${\Gamma}_{d}$ depends on the nature of final state particles and their couplings to the inflaton.  For example, if the inflaton has a Yukawa coupling $h$ to (practically) massless spin-${1 \over 2}$ fermions ${\Gamma}_{d} = {{h}^{2} \over 8 \pi}{m}_{I}$; while for the inflaton with gravitationally suppressed coupling to matter ${\Gamma}_{d} \sim {{{m}_{I}}^{3} \over {{M}_{Pl}}^{2}}$.  Efficient inflaton decay happens at ${t}_{d} = {2 \over 3{H}_{d}}$ when $H \simeq {\Gamma}_{d}$.  At this time the bulk of the energy density in the coherent oscillations of the inflaton is transferred to relativistic particles with an energy of order ${m}_{I}$.  From then on the universe is radiation-dominated, and the energy of relativistic particles is redshifted as ${a}^{-1} \propto {t}^{-{1 \over 2}}$; where $a$ is the scale factor of the universe.  After the thermalization of inflaton decay products is completed the familiar hot big-bang universe is restored.

However, inflaton decay does not suddenly happen at ${t}_{d}$.  It is rather a prolonged process which starts once the inflaton oscillations start at $t = {t}_{I}$, when $H \simeq {H}_{I}$ (${H}_{I}$ is the Hubble constant at the end of inflation).  The comoving number density of zero-mode inflaton quanta at time $t$, ${n}_{c}(t)$, obeys the relation ${n}_{c}(t) = {n}_{I} \exp({-{\Gamma}_{d}t})$; where ${n}_{I}$ is the inflaton number density at ${t}_{I}$.  The decay products have their largest number density at the earliest times, but they constitute only a tiny fraction of the energy density of the universe for $t < {t}_{d}$.  The inflaton decay becomes efficient at the time ${t}_{d}$, and at this time the bulk of the energy density is carried by the decay products.  Recently it has been noted that consideration of those decay products which were produced before ${t}_{d}$ can have important implications for thermalization \cite{mcdonald}, gravitino production \cite{grt2}, Affleck-Dine baryogenesis \cite{ace}, and electroweak baryogenesis \cite{dlr}.
    
For ${t}_{I} \leq t \leq{t}_{d}$ the universe is matter-dominated and hence $H = {2 \over 3t}$.  In the time interval between $t$ and $t + {2 \over 3}{H}^{-1} = 2t$, inflaton decay products are produced and their physical number density $n$ at the end of this interval is 

\beq
n = {n}_{I} [\exp{(-{\Gamma}_{d}t)}  - \exp{(-2{\Gamma}_{d}t)}] {({H \over 2{H}_{I}})}^{2}.
\eeq              

Here we have used the fact that for a matter-dominated universe the redshift factor for the number density of particles is ${({a \over {a}_{I}})}^{-3} = {({H \over 2{H}_{I}})}^{2}$ (the Hubble constant at the end of interval is ${H \over 2}$); where ${a}_{I}$ is the scale factor of the universe at the end of inflation.  The redshift factor for the momentum of a particle (equivalently the energy of a relativistic particle) is ${({a \over {a}_{I}})}^{-1} = {({H \over 2{H}_{I}})}^{2 \over 3}$.  For $t \leq {{t}_{d} \over 2}$ the above expression becomes $n \simeq {3H{\Gamma}_{d} \over 8{{H}_{I}}^{2}} {n}_{I}$.  The energy of these particles is between ${({1 \over 2})}^{2 \over 3}{m}_{I}$ (for the particles produced at the beginning of interval) and ${m}_{I}$ (for those produced at the end of interval), and can be practically taken as ${m}_{I}$ for all particles.  These particles have the highest energy in the spectrum of inflaton decay products at time $2t$.  On the other hand, particles which were produced in the interval $[{t}_{I},2{t}_{I}]$ have the lowest energy in the spectrum.  At $2t$, their energy has been redshifted (from ${m}_{I}$) to ${({H \over {H}_{I}})}^{2 \over 3} {m}_{I}$, while their number density has been redshifted from ${3{\Gamma}_{d} \over {H}_{I}}{n}_{I}$ to ${3{H}^{2}{\Gamma}_{d} \over 8{{H}_{I}}^{2}}{n}_{I}$.  

After changing $2t$ back to $t$, and hence $H$ to $2H$, we find that for $t < {t}_{d}$ the plasma of inflaton decay products consists of particles with energy $E$ and number density $n$

\beq
\matrix{
{({H \over {H}_{I}})}^{2 \over 3} {m}_{I} \leq E \leq {m}_{I} \cr
 \cr
 {3{H}^{2}{\Gamma}_{d} \over 4{{H}_{I}}^{3}} {n}_{I} \leq n \leq {3H{\Gamma}_{d} \over 4{{H}_{I}}^{2}} {n}_{I} \cr
}
\eeq
such that

\beq
n = {({E \over {m}_{I}})}^{3 \over 2} {3H{\Gamma}_{d} \over 4{{H}_{I}}^{2}} {n}_{I}
\eeq

Particles with energy ${E}_{Max} = {m}_{I}$ and number density ${n}_{Max} = {3H{\Gamma}_{d} \over 4{{H}_{I}}^{2}} {n}_{I}$ are produced until ${t}_{d}$.  At this time, the inflaton decay is effectively completed, and almost all of the energy density is carried by relativistic particles.  This implies that particles with energy ${m}_{I}$ are produced in the interval $[{{t}_{d} \over 2},{t}_{d}]$ for (practically) the last time.  For $H = {\Gamma}_{d}$, (2) gives the spectrum of inflaton decay products at ${t}_{d}$:

\beq
\matrix{
{({{\Gamma}_{d} \over {H}_{I}})}^{2 \over 3} {m}_{I} \leq E \leq {m}_{I} \cr
 \cr
{3{{\Gamma}_{d}}^{3} \over 2{{H}_{I}}^{3}} {n}_{I} \leq n \leq  {3{{\Gamma}_{d}}^{2} \over 4{{H}_{I}}^{2}} {n}_{I} \cr
}
\eeq

From then on the universe is radiation-dominated, (practically) no more particles with energy ${m}_{I}$ are produced by inflaton decay, and the energy and the number density of all particles will be redshifted as ${t}^{-{1 \over 2}}$ and ${t}^{-{3 \over 2}}$ respectively.  The spectrum of inflaton decay products for $H < {\Gamma}_{d}$ is

\beq
\matrix{
{{({H \over {\Gamma}_{d}})}^{1 \over 2}}{({2{\Gamma}_{d} \over {H}_{I}})}^{2 \over 3} {m}_{I} \leq E \leq {{({H \over {\Gamma}_{d}})}^{1 \over 2}}{m}_{I} \cr
 \cr
{{({H \over {\Gamma}_{d}})}^{3 \over 2}}{3{{\Gamma}_{d}}^{2} \over 2{{H}_{I}}^{2}}{n}_{I} \leq n \leq  {{({H \over {\Gamma}_{d}})}^{3 \over 2}}{3{{\Gamma}_{d}}^{2} \over 4{{H}_{I}}^{2}} {n}_{I} \cr
}
\eeq
and we also have

\beq
n = {{({H \over {\Gamma}_{d}})}^{3 \over 4}}{({E \over {m}_{I}})}^{3 \over 2} {3{{\Gamma}_{d}}^{2} \over 4{{H}_{I}}^{2}} {n}_{I}
\eeq

Let us also find the occupation number of particles, ${f}_{E}$, as a function of their energy $E$ in the spectrum.  It is seen from (2) and (5) that particles with energy $E$ at time $t$ were produced during a short interval $\bigtriangleup t$, at the time when the Hubble constant was ${H}_{p} = {({{m}_{I} \over E})}^{3 \over 2}$.  The particle momenta at the time of production were in the range ${m}_{I} - {m}_{I} {H}_{p} \bigtriangleup t \leq p \leq {m}_{I}$, and their number density at that time, from (1), was $\bigtriangleup n \simeq {n}_{I} {\Gamma}_{d} \bigtriangleup t {({H \over {H}_{p}})}^{2}$.  This results in ${f}_{E} \propto {E}^{-{3 \over 2}}$ (recall that $f \simeq {\bigtriangleup n \over {p}^{2} \bigtriangleup p}$ and here $\bigtriangleup p = {m}_{I}{H}_{p} \bigtriangleup t$).  It is clear that ${f}_{E}$ does not change in time since $\bigtriangleup n$ and ${p}^{2} \bigtriangleup p$ are both redshifted as ${a}^{-3}$. 


\section{Interactions in the Plasma of Inflaton Decay Products}

It is clear from (3) and (6) that the number density and the energy density of the plasma is completely dominated by particles with the highest energy ${E}_{Max}$ in the spectrum, both before and after efficient decay of the inflaton.  Particles with lower energy have a considerably smaller number density and hence carry a much smaller energy density.  It has been pointed out however, that lower energy particles in the spectrum may have an important role in thermalization \cite{mcdonald}.  

Thermalization is a process during which the energy density $\rho$ of a distribution of particles remains constant, while their number density $n$ changes in such a way that the mean energy of particles reaches its equilibrium value $T$.  For a distribution of relativistic particles which consists of ${n}_{B}$ bosonic degrees of freedom and ${n}_{F}$ fermionic degrees of freedom at thermal equilibrium, we have $\rho = {{\pi}^{2} \over 30} ({n}_{B} + {7 \over 8}{n}_{F}) {T}^{4}$ and $n = {\zeta(3) \over {\pi}^{2}} ({n}_{B} + {3 \over 4}{n}_{F}) {T}^{3}$.   Therefore, the ratios ${{\rho}^{1 \over 4} \over E}$ and ${{n}^{1 \over 3} \over E}$ determine the deviation from thermal equilibrium.  If these ratios are greater than $O(1)$, the number density of particles should decrease, and hence the mean energy increases in order to achieve thermal equilibrium.  If they are less than $O(1)$, the number density should increase and the mean energy will decrease.  In both cases, interactions which change the number of particles should be at equilibrium. 

Under the assumption that particles in the plasma decay very rapidly after scattering, thermal equilibrium is achieved when particle scatterings are efficient.  If only scattering of particles with energy ${E}_{Max}$ (which also have the highest number density ${n}_{Max}$) is considered, the thermalization rate is

\beq
{\Gamma}_{T} \sim {\alpha}^{2}{{n}_{Max} \over {{E}_{Max}}^{2}}
\eeq
where $\alpha$ is the gauge fine structure constant (an $O({10}^{-2})$ number).  After substituting for ${E}_{Max}$ and ${n}_{Max}$ from (5), we have

\beq
{\Gamma}_{T} \sim {\alpha}^{2}{({H \over {\Gamma}_{d}})}^{1 \over 2}{{{\Gamma}_{d}}^{2} \over {{H}_{I}}^{2}{{m}_{I}}^{2}}{n}_{I}
\eeq
Thermalization occurs when ${\Gamma}_{T} \simeq H$ and is substantially delayed in general, resulting in a low reheat temperature which is consistent with the nucleosynthesis bound on the abundance of thermally produced gravitinos \cite{eeno}.

The scattering rate of particles with energy ${E}_{s}$ and number density ${n}_{s}$ off each other is $\sim {\alpha}^{2} {{n}_{s} \over {{E}_{s}}^{2}}$.  From (6) we have ${n \over {E}^{2}} \propto {E}^{-{1 \over 2}}$, which implies that particles in a seed with energy ${E}_{s} \ll {E}_{Max}$ are scattered off each other at a much higher rate and, therefore the seed could be thermalized much earlier than the bulk of the plasma.  This has led to the notion of catalyzed thermalization of the bulk by a thermalized seed \cite{mcdonald} which happens if the number density of particles in the seed substantially increases after thermalization, and if particles in the bulk are rapidly scattered off these low energy particles.  

We have to bear in mind that $2 \rightarrow 2$ scatterings do not change the number of particles and only redistribute the energy among the scattered particles.  In order to increase the number density of particles in the plasma, the rate for one particle decay and/or $2 \rightarrow N$ scatterings (with $N \geq 3$) should be at equilibrium.  For a better analysis of thermalization it is therefore necessary to identify the relevant interactions and compare their rates.  Here we list the important interactions of a particle with mass $m$ and energy ${E}_{s}$ (assuming ${E}_{s} \gg m$, a brief note on this will come later): 

\noindent
1- $2 \rightarrow 2$ scatterings off other particles in the plasma.  It is seen from (3) and (6) that $n \propto {E}^{3 \over 2}$ and hence particles with less energy have also smaller number density and energy density.  This implies that the energy density of a seed with energy ${E}_{s}$ can considerably change only by scattering off particles with energy $E > {E}_{s}$.  For this to happen, it is also necessary that energy is redistributed among the scattred particles.  Therefore we estimate the rate for scattering of a particle with energy ${E}_{s}$ off particles with energy $E > {E}_{s}$ and number density $n$, such that the transferred energy is $\bigtriangleup E > {E}_{s}$.  The cross-section for this process is $\sigma \sim {{\alpha}^{2} \over {E}_{s}E}$ \footnote{Such scatterings occur as a result of effective contact interactions, $t$-channel processes or $s$-channel processes.  Cross-section for the desired process, i.e. scatterings with energy transfer $\bigtriangleup E > {E}_{s}$, is readily found to be $\sigma \sim {{\alpha}^{2} \over {E}_{s}E}$ for a contact interaction (e.g., the quartic coupling of bosons coming from the $D$-term part of the action in supersymmetric theories).  If $E \gg {E}_{s}$, the cross-section for the $t$-channel processes with energy transfer ${E}_{s} < \bigtriangleup E \ll E$ is much greater than ${{\alpha}^{2} \over {E}_{s}E}$.  However, these peocesses correspond to long range forces and are effectively screened in the plasma.  Overall, $\sigma \sim {{\alpha}^{2} \over {E}_{s}E}$ is a reasonable estimate when all processes are taken into account.}, and the resulting scattering rate will be of order ${\alpha}^{2} {n \over {E}_{s}E}$.  From (3) and (6) we have ${n \over E} \propto {E}^{1 \over 2}$, which implies that the highest scattering rate is off particles with energy ${E}_{Max}$ in he spectrum.   

Therefore the rate for $2 \rightarrow 2$ scatterings which increase the energy of the particle ${E}_{s}$ is ${\Gamma}_{scatt} \sim {\alpha}^{2}{{n}_{Max} \over {E}_{s}{E}_{Max}}$ \footnote{Since plasma contains particles of all energies in the spectrum we must verify that the inverse scattering is not important.  In the perturbative regime of inflaton decay ${f}_{E} <1$, and the difference ${f}_{{E}_{s}} {f}_{{E}_{Max}} - {f}_{{E}_{s} + \bigtriangleup E} {f}_{{E}_{Max} - \bigtriangleup E}$ determines the direction in which the $2 \rightarrow 2$ scatterings proceed.  As we found earlier ${f}_{E} \propto {E}^{-{3 \over 2}}$, which implies that scattering dominates over inverse scattering.}.  If ${\Gamma}_{scatt} \geq H$, particle will acquire an energy much greater than ${E}_{s}$.  However, the energy of the scatterers remains almost unchanged because their number density dominates over the other particles in the plasma.  It is seen from (2) and (5) that ${{\Gamma}_{scatt} \over H} \propto {t}^{1 \over 3}$ (${t}^{1 \over 2}$) before (after) efficient inflaton decay.  Therefore if other interactions have negligible rates (compared with scatterings), all particles will finally have energies of order ${E}_{Max}$.

\noindent
2- Decay to other particles through kinematically accessible channels with the rate ${\Gamma}_{decay} \sim \alpha {{m}^{2} \over {E}_{s}}$ (${m \over {E}_{s}}$ is the time-dilation factor).  Such decays increase the number of particles and trigger a chain of cascade decays when ${\Gamma}_{decay} \geq H$.  If ${m}^{2} < {{n}_{Max} \over {E}_{Max}}$ scatterings in (a) increase the energy of the particle before it can decay.  These scatterings will also result in a plasma-induced mass-squared of order $\alpha {{n}_{Max} \over {E}_{Max}}$
\footnote{Actually ${\alpha}^{2}{n \over E}$ is integrtaed over the whole spectrum.  It is seen from (3) and (6) that the contribution from particles with energy ${E}_{Max}$ dominates the integral.}.  Therefore ${m}^{2}$ is at least of order $\alpha {{n}_{Max} \over {E}_{Max}}$, even if the mass of the particle in the absence of plasma effects ${m}_{0}$ is very small.  As long as ${{m}_{0}}^{2} < \alpha {{n}_{Max} \over {E}_{Max}}$ we have ${\Gamma}_{decay} {\ \lower-1.2pt\vbox{\hbox{\rlap{$<$}\lower5pt\vbox{\hbox{$\sim$}}}}\ } {\Gamma}_{scatt}$, and $2 \rightarrow 2$ scatterings dominate decays.    

\noindent
3- $2 \rightarrow N$ scatterings off particles with energy ${E}_{Max}$ and number density ${n}_{Max}$, which happen at the rate ${\Gamma}_{2 \rightarrow N} {\ \lower-1.2pt\vbox{\hbox{\rlap{$<$}\lower5pt\vbox{\hbox{$\sim$}}}}\ } {\alpha}^{3} {{n}_{Max} \over {E}_{Max}{E}_{s}}$
\footnote{The extra factor of $\alpha$ appears because at least one more vertex with gauge strength interaction is needed, and there are more ${1 \over 2 \pi}$ phase space factors from extra particles in the final state.}.  These scatterings increase the number of particles too.  However, their rate is hierarchically smaller than those of $2 \rightarrow 2$ scatterings and decays, and can be neglected in the analysis below.
 

\section{Thermalization and Nucleosynthesis Bound on the Reheat Temperature}                             

The above arguments establish that the competition between $2 \rightarrow 2$ scatterings and (cascade) decays determine the thermalization time and thus the reheat temperature.  Depending on the value of ${m}_{0}$ and parameters of the model for inflation ${n}_{I}$, ${m}_{I}$, and ${\Gamma}_{d}$, different situations may arise.  Our aim here is to study these different possibilities in detail.

\noindent
1- Consider the case where ${{m}_{0}}^{2} < \alpha {{n}_{Max} \over {E}_{Max}}$ until very late times.  In this case a particle with energy ${E}_{s} $ is scattered off particles with energy ${E}_{Max}$ before it can decay.  Such scatterings increase the energy of the particle by an amount much greater than ${E}_{s}$.  This increase lowers the particle decay rate (recall that ${\Gamma}_{decay} \propto {{E}_{s}}^{-1}$) and renders the decay inefficient.  

The scatterings become important when

\beq
{\alpha}^{2} {{n}_{Max} \over {E}_{min}{E}_{Max}} \geq H
\eeq
At this time particles with energy ${E}_{min}$ are efficiently scattered and acquire much higher energies.  Subsequently the rate for scattering of particles with successively higher energies comes to equilibrium, and they acquire much higher energies as time increases.  All particles in the plasma have an energy of order ${E}_{Max}$ when

\beq
{\alpha}^{2} {{n}_{Max} \over {{E}_{Max}}^{2}} \geq H
\eeq
Shortly after that decay (now with the rate ${\Gamma}_{decay} \sim {\alpha}^{2} {{n}_{Max} \over {{E}_{Max}}^{2}}$) becomes effective, and a chain of cascade decays will lead to thermalization of the plasma.  This yields ${H}_{th} \sim {\alpha}^{2} {{n}_{Max} \over {{E}_{Max}}^{2}}$, which after replacing for ${E}_{Max}$ and ${n}_{Max}$ from (5), becomes

\beq
{H}_{th} \sim {{\alpha}^{4}{{\Gamma}_{d}}^{3} \over {{H}_{I}}^{4}{{m}_{I}}^{4}}{{n}_{I}}^{2}
\eeq
and the reheat temperature is

\beq
{T}_{R} \sim {({{\alpha}^{4}{{\Gamma}_{d}}^{3}{M}_{Pl} \over {{H}_{I}}^{4}{{m}_{I}}^{4}}}{{n}_{I}^{2})}^{1 \over 2}
\eeq
This holds if ${{m}_{0}}^{2} < \alpha {{n}_{Max} \over {E}_{Max}}$ for at least $H \geq {H}_{th}$, giving

\beq
{{m}_{0}}^{2} \leq {{\alpha}^{5}{{\Gamma}_{d}}^{4} \over {{H}_{I}}^{6}{{m}_{I}}^{5}}{{n}_{I}}^{3}
\eeq             

Therefore if the typical mass scale of particles in the plasma of inflaton decay products satisfy (13), the reheat temperature is given by (12).

\noindent
2- Consider the case where ${m}_{0}$ exceeds the bound in (14).  In this case ${{m}_{0}}^{2}$ catches up with $\alpha {{n}_{Max} \over {E}_{Max}}$ at an earlier time, and ${m}^{2} \simeq {{m}_{0}}^{2}$ subsequently.  This happens for (again we use (5))

\beq
{H}_{eq} \sim {{{H}_{I}}^{2}{m}_{I} \over \alpha {\Gamma}_{d}{n}_{I}} {{m}_{0}}^{2}
\eeq     
At this time ${\Gamma}_{scatt} \sim {\alpha}^{2} {{n}_{Max} \over {E}_{s}{E}_{Max}}$ is at equilibrium for 

\beq
{E}_{s} \sim {{\alpha}^{2}{\Gamma}_{d} \over {{H}_{I}}^{2}{m}_{I}}{n}_{I}
\eeq
Particles with an (initial) energy less than ${E}_{s}$ in the spectrum have already had efficient scatterings, and therefore have acquired much higher energies.  For particles with energies greater than ${E}_{s}$ decay dominates over scattering, and is efficient when ${\Gamma}_{decay} \sim \alpha {{{m}_{0}}^{2} \over {E}_{s}} \geq H$.  

At $H \simeq {H}_{eq}$ particles with energy ${E}_{s}$, given by (15), decay efficiently and a seed is self-thermalized after a chain of cascade decays
\footnote{Since the energy of decay products is smaller than the energy of the original particles subsequent decays in the chain occur in an even faster rate (recall that the decay rate is proportional to ${E}^{-1}$).  On the other hand, ${\Gamma}_{decay} > {\Gamma}_{scatt}$ regardless of the energy of the particle.  Therefore scatterings are not important throughout the chain.  At the end of the chain the number density of particles has increased while their energy has decreased.  This implies that particles scatter off each other very efficiently, and therefore this can be defined as the time when the seed is thermalized.}.  The energy and the number density of particles in the thermalized seed are $T \sim {{\rho}_{s}}^{1 \over 4} \simeq {({E}_{s}{n}_{s})}^{1 \over 4}$ and ${T}^{3} \sim {{\rho}_{s}}^{3 \over 4} \simeq  {({E}_{s}{n}_{s})}^{3 \over 4}$ respectively, and ${n}_{s} \ll {T}^{3} < {n}_{Max}$.  Two processes are now competing with each other: particles with an energy just above ${E}_{s}$ in the spectrum scatter off low energy particles in the thermalized seed at rate ${\alpha}^{2}{{T}^{3} \over {E}_{s}T}$, and particles in the thermalized seed scatter off particles with energy ${E}_{Max}$ at rate ${\alpha}^{2}{{n}_{Max} \over T{E}_{Max}}$.  If the first process is more efficient, particles in the bulk loose energy through scatterings off the thermalized seed.  This increases their decay rate, triggers cascade decay, and leads to their thermalization.  The thermalized seed grows very rapidly in this way until all of the bulk is thermalized.  This is called the catalyzed thermalization of the bulk by a thermalized seed \cite{mcdonald}.  If the second process is more efficient, particles in the thermalized seed acquire energies of order ${E}_{Max}$, and therefore later scatterings off the seed cannot  lower the energy of particles in the bulk.  In this case, cascade decay of particles with energies greater than ${E}_{s}$ (hence their thermalization) starts after the Hubble expansion has sufficiently redshifted their energy.  Therefore the thermalized seed grows slowly until the first process catches up with the second one.  At this time catalyzed thermalization occurs and the bulk becomes thermalized very rapidly.

For the first process to be efficient two conditions are necessary.  First, its rate should be at equilibrium

\beq
{\alpha}^{2}{{T}^{3} \over {E}_{s}T} \geq H
\eeq
and, second, it should dominate the second process

\beq
{\alpha}^{2}{{T}^{3} \over {E}_{s}T} \geq {\alpha}^{2}{{n}_{Max} \over T{E}_{Max}}
\eeq
where $T={({n}_{s}{E}_{s})}^{1 \over 4}$
\footnote{It is seen that (17) is satisfied when ${T}^{3} \geq {{E}_{s} \over {E}_{Max}}{n}_{Max}$.  This implies that catalyzed thermalization can occur when ${T}^{3} \ll {n}_{Max}$, and justifies the relation ${n}_{s} \ll {T}^{3} < {n}_{Max}$ in above.}.  After substituting for ${E}_{Max}$, ${n}_{Max}$, ${E}_{s}$, and ${n}_{s}$ from (5) and (6), and using the fact that the seed becomes thermalized after a chain of cascade decays (i.e. when $\alpha {{{m}_{0}}^{2} \over {E}_{s}} \simeq H$), (16) yields

\beq
H \leq {({{\alpha}^{18}{{\Gamma}_{d}}^{5}{{m}_{0}}^{4}{{n}_{I}}^{4} \over {{m}_{I}}^{6}{{H}_{I}}^{6}})}^{1 \over 7}
\eeq    
and (17) gives

\beq
H \leq {({{\alpha}^{2}{{m}_{0}}^{28}{{H}_{I}}^{8} \over {\Gamma}_{d}{{m}_{I}}^{2}{{n}_{I}}^{4}})}^{1 \over 21}
\eeq
Thermalization of the bulk occurs at the smaller $H$ from (19) and (20), and for $H \leq {H}_{eq}$ (the seed cannot be thermalized for $H > {H}_{eq}$ because scattering dominates over decay then)
\footnote{In reality catalyzed thermalization of the bulk occurs after a number of efficient scatterings off the thermalized seed.  It would be safer then to have a factor of 10 on the right-hand side of (16) and (17).  However, since ${T}_{R} \propto {{H}_{th}}^{1 \over 2}$, ${T}_{R}$ remains within the same order of magnitude even if ${H}_{th}$ changes by a factor of 10.}:

\beq
{H}_{th} = min~[{H}_{eq},~ {({{\alpha}^{18}{{\Gamma}_{d}}^{5}{{m}_{0}}^{4}{{n}_{I}}^{4} \over {{m}_{I}}^{6}{{H}_{I}}^{6}})}^{1 \over 7},~ {({{\alpha}^{2}{{m}_{0}}^{28}{{H}_{I}}^{8} \over {\Gamma}_{d}{{m}_{I}}^{2}{{n}_{I}}^{4}})}^{1 \over 21}]
\eeq
This holds if ${m}_{0}$ is small enough such that ${{m}_{0}}^{2} < \alpha {{n}_{Max} \over {E}_{Max}}$ before efficient inflaton decay, which, after using (4), gives

\beq
{{\alpha}^{5}{{\Gamma}_{d}}^{4} \over {{H}_{I}}^{6}{{m}_{I}}^{5}}{{n}_{I}}^{3} < {{m}_{0}}^{2} < {\alpha {{\Gamma}_{d}}^{2} \over {{H}_{I}}^{2}{m}_{I}}{n}_{I}
\eeq        

Therefore if ${m}_{0}$ satisfies (21), ${H}_{th}$ is determined from (20) and ${T}_{R} \sim {({H}_{th}{M}_{Pl})}^{1 \over 2}$.
   
\noindent
3- Consider the case where ${m}_{0}$ exceeds the bound  in (21).  In this case ${{m}_{0}}^{2}$ dominates over $\alpha {{n}_{Max} \over {E}_{Max}}$ before efficient inflaton decay (i.e., for $H \geq {\Gamma}_{d}$).  In order to find ${H}_{th}$, the above steps are repeated but the values of ${E}_{Max}$, ${n}_{Max}$, ${E}_{s}$, and ${n}_{s}$ are replaced from (2) and (3), instead of (5) and (6).  This leads to

\beq
{H}_{eq} \sim {{{H}_{I}}^{2}{m}_{I}{{m}_{0}}^{2} \over \alpha {\Gamma}_{d}{n}_{I}}
\eeq
The conditions for catalyzed thermalization of the bulk give
  
\beq
H \leq {({{\alpha}^{9}{{\Gamma}_{d}}^{2}{{m}_{0}}^{2}{{n}_{I}}^{2} \over {{m}_{I}}^{3}{{H}_{I}}^{4}})}^{1 \over 3}
\eeq
instead of (18), and

\beq
H \leq {({{\alpha}^{7}{{H}_{I}}^{4}{{m}_{0}}^{14} \over {m}_{I}{{\Gamma}_{d}}^{2}{{n}_{I}}^{4}})}^{1 \over 9}
\eeq
instead of (19).  Therefore in this case

\beq
{H}_{th} = min~[{({{\alpha}^{9}{{\Gamma}_{d}}^{2}{{m}_{0}}^{2}{{n}_{I}}^{2} \over {{m}_{I}}^{3}{{H}_{I}}^{4}})}^{1 \over 3},~{({{\alpha}^{7}{{H}_{I}}^{4}{{m}_{0}}^{14} \over {m}_{I}{{\Gamma}_{d}}^{2}{{n}_{I}}^{4}})}^{1 \over 9}]
\eeq
It is also necessary that ${H}_{th} \geq {\Gamma}_{d}$ for thermalization to occur before efficient decay of the inflaton (otherwise we are back to  case 2 in the above) which implies that ${m}_{0}$ must be indeed large.  In this case the plasma is thermal when inflaton decay is completed at $H \simeq {\Gamma}_{d}$ and ${T}_{R} \sim {({\Gamma}_{d}{M}_{Pl})}^{1 \over 2}$.

In summary, depending on the value of ${m}_{0}$ and inflationary model parameters ${n}_{I}$, ${m}_{I}$, and ${\Gamma}_{d}$, we have

\beq 
{({{\alpha}^{4}{{\Gamma}_{d}}^{3}{M}_{Pl} \over {{H}_{I}}^{4}{{m}_{I}}^{4}}}{{n}_{I}^{2})}^{1 \over 2} {\ \lower-1.2pt\vbox{\hbox{\rlap{$<$}\lower5pt\vbox{\hbox{$\sim$}}}}\ } {T}_{R} {\ \lower-1.2pt\vbox{\hbox{\rlap{$<$}\lower5pt\vbox{\hbox{$\sim$}}}}\ } {({\Gamma}_{d}{M}_{Pl})}^{1 \over 2}
\eeq
Big-bang nucleosynthesis yields the bound ${T}_{R} {\ \lower-1.2pt\vbox{\hbox{\rlap{$<$}\lower5pt\vbox{\hbox{$\sim$}}}}\ } {10}^{8}-{10}^{10}$ GeV, which constrains the parameters of the model.  This ensures that gravitinos are not produced in dangerous abundances in the thermal bulk.  

Even before thermalization of the bulk gravitinos are produced in the non-thermal bulk and the thermal seed.  One may wonder whether the requirement ${T}_{R} {\ \lower-1.2pt\vbox{\hbox{\rlap{$<$}\lower5pt\vbox{\hbox{$\sim$}}}}\ } {10}^{8}-{10}^{10}$ GeV also keeps the abundance of these gravitinos at a safe level.  The number density of gravitinos which are produced in the plasma is in general proportional to ${n}^{2}$ ($n$ is the typical number density of particles in the plasma)
\footnote{One factor of $n$ comes from the scattering rate and the other one is from the number density of particles in the plasma.}, which is much smaller before thermalization of the bulk.  On the other hand, thermalization of the bulk releases a huge entropy and dilutes everything, including gravitinos which were produced in the thermal seed.  In the case that thermalization occurs before efficient inflaton decay (case 3 in the above) it has been shown that the bound ${T}_{R} {\ \lower-1.2pt\vbox{\hbox{\rlap{$<$}\lower5pt\vbox{\hbox{$\sim$}}}}\ } {10}^{8}-{10}^{10}$ GeV guarantees successful nucleosynthesis, despite the fact that the instantaneous temperature of the plasma is much higher at earlier times \cite{grt2}.  We can expect that the same conclusion also holds for cases 1 and 2 above and therefore the strongest constraint on the model parameters is, in general, due to gravitino production in the thermal bulk.      

Two clarifying comments are to be made here.  We have actually studied the thermalization of particles with mass ${m}_{0}$.  After their thermalization, the lighter particles reach thermal equilibrium through gauge-strength scatterings of these particles (e.g., two squarks exchange a gaugino and scatter to two quarks).  Therefore only the typical mass scale of the sector is important for its thermalization and not the individual particle masses.  Finally, the above analysis was done for relativistic particles and its consistency requires that ${T}_{R} \geq {m}_{0}$.  If ${T}_{R} < {m}_{0}$, particles with mass ${m}_{0}$ become non-relativistic before thermalization occurs.  These particles will then decay to other (relativistic) particles whose subsequent (cascade) decay leads to thermalization and determines the reheat temperature.  The same analysis as above can be done in this case, but ${m}_{I}$ and ${m}_{0}$ should be replaced by ${m}_{0}$ and the mass scale of secondary decay products respectively.      


\section{The MSSM Example and the Role of Flat Directions}
 
Here we consider a model which consists of the Minimal Supersymmetric Standard Model (MSSM) sector, the inflaton sector, and the low-energy supersymmetry breaking sector, where the latter two interact with the MSSM sector through gravitationally suppressed couplings.  To keep the matter content minimal, no right-handed neutrinos or gauge sectors at the intermediate scale are introduced, and therefore the MSSM sector is the only one with gauge strength couplings.  We consider a chaotic inflation model with the potential ${1 \over 2}{{m}_{I}}^{2}{\phi}^{2}$ for the inflaton $\phi$, where the inflaton decays to other particles via gravitationally suppressed couplings (and hence there is no stage of parametric resonance decay in this model).  From the COBE data on the isotropy of the microwave background radiation, ${m}_{I} \simeq {10}^{13}$ GeV \cite{cdo2}.  At the end of inflation $\phi \simeq {10}^{19}$ GeV which yields 

\begin{eqnarray*}
{H}_{I} \simeq {{m}_{I}\phi \over {M}_{Pl}} \simeq {10}^{13}~{\rm GeV}\\
{n}_{I} = {m}_{I} {\phi}^{2} \simeq {10}^{51}~ {\rm GeV}^{3}\\
{\Gamma}_{d} \sim{{{m}_{I}}^{3} \over {{M}_{Pl}}^{2}} \simeq 10~ {\rm GeV}\\
\end{eqnarray*}
Also for the MSSM sector, ${m}_{0} \simeq {10}^{2}$ GeV.  It is easily seen that with these values, (13) is satisfied and the reheat temperature derived from (12) is ${T}_{R} \sim {10}^{6}$ GeV, which is far below the limit set by the nucleosynthesis bound.

In the early universe however, it is possible to have much larger mass scales in the MSSM sector.  Flat directions in the scalar potential of the MSSM (denoted as $\varphi$ here) can acquire very large vevs during inflation, both in the minimal models \cite{drt} and in the no-scale models \cite{gmo} of supergravity.  At the end of inflation $\varphi \simeq {({H}_{I}{M}^{n-3})}^{1 \over n-2}$ \cite{drt}, where $n$ (not to be confused with the number density) is the order of nonrenormalizable superpotential term which lifts the flat direction, and $M$ is the scale of new physics which induces the nonrenormalizable terms.  The non-zero energy density of the early universe strongly breaks supersymmetry and induces a negative mass-squared of order $-{H}^{2}$ for the flat direction.  So long as other supersymmetry breaking sources do not induce a positive mass-squared of the same order, $\varphi \simeq {(H{M}^{n-3})}^{1 \over n-2}$ and the flat direction vev decreases slowly.  When other sources become dominant, the flat direction mass-squared becomes positive, oscillations start,  and from then on $\varphi$ is Hubble redshifted more rapidly.  

A non-zero vev for the flat direction breaks a subgroup of the MSSM gauge group.  Those MSSM fields which have $F$- and $D$-term couplings to the flat direction acquire masses of order $g \varphi$, where $g$ is a gauge or Yukawa coupling.  This implies that in the early universe the mass scale of those fermions and scalars which are coupled to the flat direction, as well as the gauge fields and gauginos of the broken subgroup, can be much larger than ${10}^{2}$ GeV.  Thermalization in the MSSM sector can then happen earlier (case 2 or 3 in the above), leading to a reheat temperature higher than ${10}^{6}$ GeV.  For example, if ${m}_{0} \sim{10}^{10}$ GeV (which can be easily induced by the flat direction vev), thermalization occurs before efficient inflaton decay and ${T}_{R} \sim {10}^{10}$ GeV.  On the other hand, the flat direction vevs, and thus the induced masses, are rapidly redshifted after the flat direction oscillations start.  The question is whether they remain large enough for a sufficiently long time so that thermal equilibrium is achieved.  This depends on the time when oscillations start and the flat direction vev at the onset of oscillations.  In the standard approach \cite{drt}, oscillations start at $H \simeq {m}_{3 \over 2} \simeq {10}^{2}-{10}^{3}$ GeV when the low-energy supersymmetry breaking takes over the Hubble-induced one.  Recently it has been shown that many flat directions can start their oscillations much earlier, due to plasma effects \cite{ace}.  Therefore the mass scale of the model at early times, and hence the thermalization dynamics, has a dependence on the nature of the flat direction and its initial vev.  This may suggest another role for the supersymmetric flat directions besides baryogenesis \cite{ad} and dark matter candidates \cite{ks}: they may also assist thermalization to happen earlier and lead to a higher reheat temperature.  

In conclusion the reheat temperature ${T}_{R}$ may be below the upper limit ${({\Gamma}_{d}{M}_{Pl})}^{1 \over 2}$ by several orders of magnitude, in which case the constraints on the parameters of the model are considerably relaxed
\footnote{For example, in models of $D$-term inflation or supersymmetric models for new inflation the inflaton mass can be be as large as ${10}^{15}-{10}^{16}$ GeV \cite{lr}.  In such models the upper limit on ${T}_{R}$ can be as high as ${10}^{12}$ GeV.  If ${m}_{0}$ is small enough,  the reheat temperature will be far below its upper limit and hence in agreement with the nucleosynthesis bound.}.  However, flat directions with large vevs may result in earlier thermalization and push  ${T}_{R}$ towards its upper limit.


\section{Conclusion}      

We have considered thermalization after perturbative decay of the inflaton.  In this regime the number density of inflaton decay products is smaller than the thermal equilibrium value, while the mean energy of particles is larger than its thermal value.  Particles with the highest energy ${E}_{Max}$ dominate the number density and the energy density of the plasma of inflaton decay products, and $2 \rightarrow 2$ scatterings off these particles increase the energy of other particles.  We compared the rate for $2 \rightarrow 2$ scatterings, particle decay, and $2 \rightarrow N$ scatterings and found that the first two are the important ones.  Thermalization occurs when decays dominate scatterings and, depending on the model parameters, different situations may arise.  We showed that the thermalization time and the reheat temperature has a dependence on the typical mass scale of inflaton decay products.  If this mass scale is sufficiently large, a seed in the plasma is self-thermalized and the scattering of particles in the plasma off the thermalized seed leads to catalyzed thermalization of the bulk.  Otherwise, the bulk is self-thermalized at late times, when all particles in the plasma have energies of order ${E}_{Max}$.  As a result, the reheat temperature can vary in a wide range.  The strongest constraint (from nucleosynthesis) on the model parameters is derived for a very large mass scale when thermalization occurs before efficient inflaton decay.  In general, a small mass scale considerably relaxes this constraint.  We also considered the MSSM case as an example.  It was shown that the mass scale of the model can be much larger in the early universe due to particle couplings to the flat directions with very large vevs.  If the flat direction oscillations do not start early, these vevs stay large enough for a sufficiently long time.  In this case, thermalization may occur earlier, thus leading to a higher reheat temperature and resulting in a stronger constraint on the model parameters.  Finally, the dynamics of thermalization seems to be very detailed even in the perturbative regime of inflaton decay.


\noindent{ {\bf Acknowledgements} } \\
\noindent 
The author wishes to thank B. A. Campbell for useful discussions and comments.  He also thanks A. N. Kamal for discussions and K. Kaminsky for careful reading of the manuscript.  This work was supported in part by the Natural Sciences and
Engineering Research Council of Canada.

\end{document}